\tolerance = 10000 \documentstyle[11pt]{article}
\newcommand{\ep}{\epsilon}
\newcommand{\be}{\begin{equation}}
\newcommand{\ee}{\end{equation}} %\indent}
\newcommand{\eei}{\end{equation}\indent\indent}
\newcommand{\bc}{\begin{center}}
\newcommand{\ec}{\end{center}}
\newcommand{\ber}{\begin{eqnarray}}
\newcommand{\ear}{\end{eqnarray}}
\newcommand{\ba}{\begin{array}}
\newcommand{\ea}{\end{array}}
\newcommand{\n}{{}^{(3)}\nabla} \newcommand{\na}{\nabla}

\newcommand{\hs}{\,-\,}

\newcommand{\3}{{}^{(3)}}

\newcommand{\dd}{{\cal D}}
\newcommand{\es}{{\cal E}}
\newcommand{\kk}{{\cal K}}
\newcommand{\ls}{{\cal L}}

\newcommand{\zz}{{\cal Z}}

\newcommand{\wt}{\widetilde}

\newcommand{\inu}{ {(\nu)} }

\newcommand{\sfrac}[2]{{\textstyle{#1\over#2}}}
\def\case#1/#2{\textstyle\frac{#1}{#2} }
\begin{document}
%%%%%%%%%%%%%%%%%%%%%%%%%%%%%%%%%%%%%%%%%%%%%%%%%%%%%%%%
\begin{titlepage}
\addtolength{\textwidth}{0.2\textwidth}
\title{\bf Conserved Quantities \\
in Perturbed Inflationary Universes }
\author{
   {\sc Peter K. S. Dunsby}\\
   \normalsize{ {\it Department of
Applied Mathematics, University of Cape Town }}\\
\normalsize{  {\it Rondebosch 7700, Cape Town, South Africa.}}\\
\normalsize{ {\it School of Mathematical Sciences, Queen Mary and Westfield
College,}} \\
\normalsize{ {\it University of London, Mile End Road,
London E1 4NS, UK.}}\\
 {\sc  Marco Bruni}\\
\normalsize{{\it School
of Mathematical Sciences, Queen Mary and Westfield  College,}} \\
\normalsize{ {\it University of London, Mile End Road,
London E1 4NS, UK.}}\\
\normalsize{\it Dipartimento di Astronomia,
Universit\`a degli Studi di Trieste,}\\
\normalsize{\it via Tiepolo 11, 34131 Trieste, Italy.} }
\date{$\mbox{}$ \vspace*{0.3truecm} \\ \normalsize  December 2 1993}
%put the date here, before }
%%%%%%%%%%%%%%%%%%%%%%%%%%%%%%%%%%%%%%%%%%%%%%%%%%%%%%%%%%%%%%%%%%%%%%
\maketitle
\thispagestyle{empty}
%\vspace*{1truecm}
\begin{abstract}
%This paper deals with gauge\hs invariant perturbations of inflationary
%Robertson\hs Walker universes.

Given that observations seem to favour a \index{density parameter}
$\Omega_0<1$, corresponding to an open universe, we consider
gauge\hs invariant perturbations of non\hs flat Robertson\hs Walker
universes filled with a general imperfect fluid which can also be
taken to represent a scalar field. Our aim is to set up the equations
that govern the evolution of the density perturbations $\Delta$ so
that it can be determined  through a {\it first order differential
equation} with a quantity $\kk$ which is conserved at any  length
scale, even in non\hs flat universe models, acting as a source term.
The quantity $\kk$ generalizes other variables that are conserved in
specific cases (for example at large scales in a flat universe) and is
useful to connect different epochs in the evolution of density
perturbations via a transfer function. We show that the problem of
finding a conserved $\kk$ can be reduced to determining two auxiliary
variables $X$ and $Y$, and illustrate the method with two simple
examples.
\vspace*{0.5truecm}
\bc Ref. SISSA 163/93/A (Dec. 93)\ec
\end{abstract} \vspace*{0.2truecm}
\bc{\it Subject headings:}\\ cosmology \hs galaxies: clustering,
formation \hs hydrodynamics \hs relativity \ec
\end{titlepage}
%\end{document}
%%%%%%%%%%%%%%%%%%%%%%%%%%%%%%%%%%%%%%%%%%%%%%%%%%%%%%%%%%%%%%%%%%%%%%
% \setlength{\baselineskip}{24pt}
%%%%%%%%%%%%%%%%%%%%%%%%%%%%%%%%%%%%%%%%%%%%%%%%%%%%%%%%%%%%%%%%%%%%%%
\section{Introduction}
%%%%%%%%%%%%%%%%%%%%%%%%%%%%%%%%%%%%%%%%%%%%%%%%%%%%%%%%%%%%%%%%%%%%%%
Despite much theoretical progress made in the last fifteen years, the
problem of large--scale structure formation in the universe remains
an  unresolved puzzle for cosmologists. In fact, as new and more
reliable data becomes available, none of the models proposed during
this time seem satisfactory.

In the standard Hot Big Bang model \cite{bi:peeblesetal}, the most
popular mechanism advocated to explain structure formation is that of
the gravitational instability: small perturbations created during an
inflationary era eventually grow in the matter dominated era.
It is therefore necessary, in order to use observational data to
constrain theoretical models, to relate the original amplitude of a
perturbation mode at an initial time to that of the same mode today,
i.e. to derive a transfer function (cf. \cite{bi:lily,bi:efst}).

In cosmological perturbation theory the evolution of a perturbation
mode is usually given by a second order ordinary differential equation
(e.g. see \cite{bi:bardeen,bi:MFB,bi:BDE,bi:lily}). It is
therefore very useful to set up the problem in such a way  that, at
least in the simplest cases, one has a conserved quantity at ones
disposal, so that the problem is effectively reduced to solving a
first order differential equation with a constant source term which, at
least in principle, can always be integrated.

Indeed, such conserved quantities can be found in flat
(see \cite{bi:BST,bi:lyth,bi:EBH} and
\cite{bi:HV,bi:jchan1,bi:BED,bi:lily}) and almost
flat\cite{bi:BE,bi:jchan1}) universe model, for a perturbation mode
whose wavelength is much larger than the Hubble radius. This is quite
satisfactory in the standard inflationary scenario \cite{bi:olive},
where the  present day density parameter $\Omega_0$ is predicted to be
equal to  unity with a high degree of accuracy. However, there is no
compelling  observational evidence for the case of $\Omega_0=1$;
rather, a low  density universe seems to be favoured
by analysis of present day observational data \cite{bi:KT}.

For this reason, and taking into account the appeal of the
inflationary paradigms and the fact that this is up to now the only
scenario in which perturbations are predicted (by quantum
fluctuations within the horizon) rather than assumed as initial data,
some authors\cite{bi:omega3,bi:ELM,bi:EM1,bi:HE} have considered the
possibility of an era of ``minimal" inflation, in which the
accelerated expansion does not last long enough to drive $\Omega_0$
to unity today. This possibility can be a source of a certain amount
of debate \cite{bi:BANF}, raising on one hand arguments of fine tuning
(which however is in one way or an another needed for other reasons in
most of the inflationary models  proposed up to now), and on the
other hand, ``how to put a  measure on the set of initial data"
(lacking this measure, according to this point of view, the fine
tuning argument is a weak one).

In this paper we do not enter in such a debate, instead we
take the case of a negatively curved universe as an open possibility,
and examine perturbations in this context. For the sake of generality,
we shall consider an imperfect fluid, which later we shall take to
represent a perturbed minimally coupled scalar field.

Our aim is to set up the equations that govern the evolution of the
density perturbation mode $\Delta^{(\nu)}$ so that it can be
determined by a first order equation with a conserved quantity
$\kk^{(\nu)}$ as a source term.
In other words, we define a simple algorithm that in principle allows
us to find a new quantity $\kk^{(\nu)}$ which is conserved {\it at any
scale}, even in non-flat models. Then $\kk$ generalizes known
conserved  quantities, and allows one to define a transfer function in
a more  general set of circumstances.

As it will be shown, the problem of finding the quantity $\kk^{(\nu)}$
in the cases of: {\it i)} adiabatic perturbations in a perfect fluid
or {\it ii)} perturbations in a minimally coupled scalar field,
reduces to determine two auxiliary functions $X$ and $Y$.

It is noticeable that a completely consistent theory of large scale
Cosmic Microwave Background (CMB) fluctuations in open models seems to
be lacking until now \cite{bi:panek}.  With the cautionary note that
even the concept of power  spectrum has not yet been generalized to
include perturbations over super horizon scales in open universe
models, we hope that the work presented here is a step towards a
consistent and useful treatment of fluctuations in $\Omega_0 <1$
universes.

This paper is organized as follows: In section 2 we discuss very briefly
the formalism and variables used. In section 2 and 3 we define two basic
curvature variables and discuss under what conditions they are conserved both
in the case of an imperfect fluid and a minimally coupled scalar field.
In section 4 we give a prescription for finding quantities that are
conserved at all scales for any given background evolution. In
section 5 we apply our algorithm in two simple cases: {\it i)} a
coasting ($\Omega=const$) phase and {\it ii)} a De-Sitter inflationary
phase. In section 6 we briefly summarize how to use our algorithm in
order to find a transfer function. Finally in section 7 we end with a
discussion.

In this paper we assume standard general relativity, with Einstein
equations in the form $R_{ab} -\sfrac12 R g_{ab} +\Lambda g_{ab}=\kappa
T_{ab}$,  $\kappa$ and $\Lambda$ being the gravitational and
cosmological  constants respectively.
%%%%%%%%%%%%%%%%%%%%%%%%%%%%%%%%%%%%%%%%%
\section{Formalism} \label{sec:prelims}
%%%%%%%%%%%%%%%%%%%%%%%%%%%%%%%%%%%%%%%%%
For the sake of self\hs consistency, we sketch in this section the
covariant approach to  density inhomogeneities in a general curved
spacetime \cite{bi:EB}.  This as been applied to almost
Friedmann\hs Lema\^{i}tre\hs Robertson\hs Walker (FLRW) universes
dominated by a barotropic fluid \cite{bi:EHB,bi:EBH}, as well as by a
simple  scalar field \cite{bi:BE,bi:BED}, a general imperfect fluid
\cite{bi:HV,bi:BDE} and a mixture of interacting fluids
\cite{bi:PKSD,bi:DBE}. Here we just give an outline of the material
which will be needed later on, with emphasis on known conserved
quantities.
%%%%%%%%%%%%%%%%%%%%%%%%%%%%%%%%%%%%%%%%%
\subsection{Covariant cosmology}
%%%%%%%%%%%%%%%%%%%%%%%%%%%%%%%%%%%%%%%%%
Let us consider an imperfect fluid flow in a general curved space\hs time.
The 4\hs velocity of the fluid is the tangent along the fluid flow:
$u^a=dx^a/d\tau$ ($u^au_a=-1$),
where $\tau$ is the proper time along the flow lines (the world lines
of observers comoving with the fluid).
Introducing  a projection tensor orthogonal to $u^a$,
\cite{bi:ellis1}
$h_{ab}=g_{ab} +u_a u_b$ ($h_a{}^b u_b =0$)
the energy\hs momentum tensor is decomposed as follows:
\be
T_{ab}=\mu u_a u_b +p h_{ab} +2q_{(a}u_{b)} +\pi_{ab}\;, \label{eq:emt}
\ee
where $\mu$ and $p$ are the energy density and pressure of the fluid,
and $q_a$ and $\pi_{ab}$ are the energy flux and anisotropic pressure
contributions. When the fluid is not perfect (i.e. in it's perturbed
state) the fluid 4-velocity $u^a$ is not
uniquely defined \cite{bi:BDE,bi:DBE}: it is however standard  to choose
$u^a$ either as the {\it particle frame}, in which $T_{ab}$ has the
general form (\ref{eq:emt}), or as the {\it energy frame},  in which
$q_a=0$ (for a discussion of these choices, see \cite{bi:BDE}). Here
we  shall take the particle frame choice, as this allows one also to describe
more general cases for example a non\hs minimally coupled scalar
field and  higher order gravity theories \cite{bi:madsen}.

The covariant derivative of $u_a$ can be split into four parts:
\be
u_{a;b}=\sfrac13\Theta h_{ab}+\sigma_{ab} +\omega_{ab} -a_a u_b\;,
\label{eq:usplit}
\ee
where $\Theta\equiv u^a{}_{;a}$ is volume the expansion,
$\sigma_{ab}=\sigma_{(ab)}$
is the shear ($\sigma_{ab}u^b=\sigma^a{}_a=0$),
$\omega_{ab}=\omega_{[ab]}$ is the vorticity ($\omega_{ab}u^b=0$) and
$a_a=\dot{u}_a=u_{a;b}u^b$ is the acceleration (the dot denotes the
proper time derivative). It is useful to introduce a {\it length}
scale factor along flow lines by the relation
\be
\frac{1}{3}\Theta=\frac{\dot{\ell}}{\ell}=H\;;
\ee
when the universe is an exact FLRW space\hs time $H$ is just the
usual Hubble parameter. The  evolution equation for the expansion
$\Theta$ is the Raychaudhuri equation
\be
\dot{\Theta} +\sfrac13\Theta^2 +2(\sigma^2-\omega^2)-a^a{}_{;a}+\sfrac12
\kappa(\mu+3p)-\Lambda=0\;, \label{eq:rai}
\ee
where $\sigma^2=\sfrac12\sigma_{ab}\sigma^{ab}$ and
$\omega^2=\sfrac12\omega_{ab} \omega^{ab}$ are the shear and vorticity
magnitudes.

The matter equations of motion $T^{ab}{}_{;b}=0$ are equivalent to the
energy conservation and  momentum conservation equations: retaining
only first order contributions in  the inhomogeneity variables (the
reason for this will be clarified in  the next section) we  have:
\be
\dot{\mu}+3hH+h\n_a\Psi^a=0\;,
\label{eq:econ}
\ee
and
\be
h a_a+Y_a+h\left[F_a+\Pi_a\right]=0\;,
\label{eq:mcon}
\ee
where for later convenience we define
\be
h=\left(\mu+p\right),~~~ \Psi_a=\frac{q_a}{h}\;,~~~ ,
F_a=\dot{\Psi}_a-\left(3c^2_s-1\right)H\Psi_a\;,~~~\Pi_a=\frac{1}{h}
\n^b\pi_{ab}\;. \label{eq:defnewnot}
\ee

An exact FLRW can be covariantly characterized by the vanishing of the
shear and the vorticity of $u^a$ and by the vanishing of the spatial
gradients (i.e. orthogonal to $u^a$) of any scalar $f$:
\be
\sigma =\omega=0\;, ~~~\n_a f =0\;; \label{eq:FLRW1}
\ee
in particular the gradients of energy density, pressure and expansion
\be
X_a\equiv\n_a\mu\;, ~~Y_a\equiv\n_a p\; ~~Z_a\equiv\n_a\Theta\;,
\label{eq:FLRW2}
\ee
vanish, where $Y_a=0\Rightarrow a_a=0$. Then $\mu=\mu(t)$, $p=p(t)$ and
$\Theta=\Theta(t)=3 H(t)$ depend  only on the cosmic time $t$
defined (up to a constant) by the FLRW fluid flow vector through
$u_a=-t_{,a}$. The energy momentum tensor (\ref{eq:emt}) necessarily
reduces to the perfect fluid form $T_{ab}=\mu u_a u_b +p h_{ab}$,
and it follows that these models are completely determined
by an equation of state $p=p(\mu)$, the energy conservation
equation (\ref{eq:econ}) (with no imperfect fluid contributions)
and the Friedmann equations
\be
3\dot{H} +3 H^2 +\sfrac12\kappa(\mu+3p) -\Lambda=0\; ,
\label{eq:FLRW4}
\ee
and
\be
H^2 +\case{K}/{\ell^2}=\sfrac13 \kappa\mu +\sfrac13\Lambda\;,
\label{eq:FLRW5}
\ee
where the latter is a first integral of the former (when $H\not= 0$),
which  is  (\ref{eq:rai}) specialized to a FLRW model.
%%%%%%%%%%%%%%%%%%%%%%%%%%%%%%%%%%%%%%%%%%%%%%%%%%%%%%%%%%%
\subsection{Fluid inhomogeneities and gauge invariant  variables}
%%%%%%%%%%%%%%%%%%%%%%%%%%%%%%%%%%%%%%%%%%%%%%%%%%%%%%%%%%%
The vector $Y_a$ is one of a set of variables we can
define \cite{bi:EB} to describe the fluid inhomogeneity:
\be
Y_a \equiv h_a{}^b p_{,b}\;,
{}~~~\dd_a\equiv \ell\frac{X_a}{\mu}\;,
{}~~~\zz_a  \equiv  \ell h_a{}^b \Theta_{,b}\;;
\ee
these are the covariantly defined
 {\it spatial} (i.e. orthogonal to $u^a$) gradients of the
pressure, energy density and expansion; more precisely the
dimensionless comoving fractional spatial gradient of the energy
density $\dd_a$ is the physically relevant variable to characterize
the energy density spatial variation. All of these covariantly defined
exact variables can be in principle measured as spatial gradients by
observers at rest with respect to the fluid,  and in general (i.e. in
an arbitrary spacetime) they obey  exact non\hs linear and covariant
evolution equations which have been  derived in \cite{bi:EB} for a
perfect fluid and in \cite{bi:PKSD} for an imperfect fluid; these
equations are coupled with the exact evolution  equations for
$\sigma_{ab}$ and $\omega_{ab}$ (see e.g. \cite{bi:ellis1}).

The above introduced covariant gradients $\zz_a$ and $\dd_a$, and in
general any variable such as those in (\ref{eq:FLRW1}) that vanish in
the FLRW background, are gauge\hs invariant (GI) variables  (for more
details, see \cite{bi:bardeen,bi:EB,bi:MFB} and \cite{bi:stewart}),
so  we can proceed to obtain equations for them valid in an {\it
almost} FLRW universe (a space\hs time where these variables are
small), approaching this  universe from a general space\hs time rather
than perturbing an exact  FLRW model. Then the linearization procedure
we apply consists in dropping terms such as $\sigma^2$ in
(\ref{eq:rai}), i.e. terms which are of second order in the GI
variables, retaining only terms linear in $\dd_a$ and $\zz_a$ and
using zero\hs order expressions for $\mu$, $p$ and $\Theta=3H$ [this
clarifies the meaning of neglecting higher order terms in
(\ref{eq:econ}) and (\ref{eq:mcon})].

At zero\hs order these variables are functions of the cosmic time only,
and satisfy the usual equations valid in an exact FLRW universe.
Applying this linearization procedure to a universe where the fluid is
barotropic [$p=p(\mu)$], one obtains a closed pair of first\hs
order (in proper time derivatives) equations that couple $\dd_a$ and
$\zz_a$ only\,\cite{bi:EB};  this is equivalent to a second\hs order
equation for $\dd_a$\,\cite{bi:EHB,bi:EBH}. It was noted in \cite{bi:EBH}
that in general a source term due to vorticity is present in this
equation; it also admits an entropy perturbation source term when a
more general non barotropic fluid is considered\,\cite{bi:HV}.
Finally, we note that our covariant exact variables can be related to
those of Bardeen at first order (see \cite{bi:BDE,bi:HV}). Using this
it is then possible to connect our GI perturbation variables to the
standard analysis of the spectrum of perturbations and fluctuations in
the cosmic microwave background, which are usually given in some
specific gauge.
%%%%%%%%%%%%%%%%%%%%%%%%%%%%%%%%%%%%%%%%%%%
\subsection{Conserved quantities} \label{sec:cons}
%%%%%%%%%%%%%%%%%%%%%%%%%%%%%%%%%%%%%%%%%%%
In order to proceed towards the identification of GI conserved
quantities, we introduce the 3\hs curvature scalar of the projected
metric  $h_{ab}$ orthogonal to $u^a$; which in general is (see the
appendix in \cite{bi:EBH})
\be
\3 R=2(-\sfrac13 \Theta^2+\sigma^2-\omega^2 +\kappa\mu+\Lambda)\;,
\label{eq:curv}
\ee
and when $\omega=0$ it reduces to the 3-curvature of the surfaces
orthogonal to  $u^a$. From (\ref{eq:curv}) we can define the GI
curvature gradient as
\be
C_a\equiv \ell^3 h_a{}^b \3 R_{,b}=-\sfrac43\Theta \ell^2\zz_a+2\kappa\mu
\ell^2 \dd_a\;, \ee
where the second equality holds in the linear approximation;
then this  expression shows that only two of the three variables
$\dd_a$, $\zz_a$ and $C_a$ are linearly independent at first  order.

Up to now we have considered vectorial GI variables. However we are
interested here in characterizing matter clumping through scalar
density perturbations: instead of introducing these through  the
standard non\hs local (ADM) splitting\,\cite{bi:stewart}, we simply
locally define GI scalar variables by taking the divergence of the
vectorial quantities. In particular we have:
\be
\Delta=\ell h^{ab}\dd_{a;b}\;, ~~~\zz=\ell h^{ab}\zz_{a;b}\;,~~~
C=-\sfrac43\Theta \ell^2\zz+2\kappa\mu \ell^2\Delta\;,
\ee
where $\Delta$ is the locally defined density perturbation scalar
variable describing matter clumping (the factor $\ell$ in these
definitions comes in for dimensional reasons). We also introduce
the related quantity
\be
\wt{C}=C-4 K\gamma^{-1}\Delta\;, ~~~\gamma=1+w\;, ~~~w=\frac{p}{\mu}\;,
\label{eq:cdef}
\ee
which reduces to the GI curvature perturbation $C$ when $K=0$ [$K$ is
the curvature constant of FLRW models appearing in (\ref{eq:FLRW5})].
Once a harmonic analysis is carried out,
$\Delta$ is equivalent to the Bardeen variable $\varepsilon_m$
\cite{bi:HV,bi:BDE} and $\wt{C}$ corresponds to the variable $\zeta$
defined in \cite{bi:BST} (see \cite{bi:jchan1}).
Again only two of our variables are independent, therefore the $\Delta$
evolution can be obtained from a linear system of two equations for $\Delta$
and $\wt{C}$ or $\Delta$ and $\zz$. For a general fluid with speed of sound
$c_s^2\equiv dp/d\mu$ we obtain\,\cite{bi:BDE}:
\ber
\lefteqn{
\dot{\Delta}-\left\{3Hw-\left[\case{\kappa
 h}/{2}-\case{K}/{\ell^2}\right]H^{-1}\right\}
\Delta-\case{(1+w)}/{4\ell^2H}\tilde{C}}\nonumber\\
&=&3\ell\left(1+w\right)H\left[F+\Pi\right]-\ell\left(1+w\right)
\na^2\Psi\;,
\label{eq:deltadot}
\ear
\ber
\dot{\zz}&+&2H\zz+\case{1}/{2}\kappa\mu\Delta+\case{c_s^2}/{\left(1+w\right)}
\left(\n^2+\case{3K}/{\ell^2}\right)\Delta
\case{w}/{(1+w)}\left(\n^2+\case{3K}/{\ell^2}\right)\es\nonumber\\
&=&-\ell\left(\n^2+\case{3K}/{\ell^2}\right)\left[F+\Pi\right]
+\case{3}/{2}h\left[F+\Pi\right]\;.
\label{eq:scalarZ}
\ear
and
\ber
\dot{\wt{C}}&=&\case{4\ell^2Hc^2_s}/{\left(1+w\right)}\n^2\Delta+
\case{4\ell^2Hw}/{(1+w)}
\left[\n^2+\case{3K}/{\ell^2}\right]\es\nonumber\\
& & +4\ell^3H\n^2\left[F+\Pi\right]+\left[4K\ell-2\ell^3h\right]\n^2\Psi\;,
\label{eq:cdot}
\ear
where
\be
\Pi=\ell\n^a\Pi_a\;, ~~~~F=\ell\n^a F_a\;, ~~~~\Psi=\ell\n^a\Psi_a\;.
\ee
The system of equations
(\ref{eq:deltadot}-\ref{eq:cdot})
admits $\dot{\gamma}\not=0$ (when $\dot{\gamma}=0$ one has $c_s^2
=w=\gamma-1$) and is derived for general $K$ and $\Lambda$;
however in the following we shall assume $\Lambda=0$.

It is useful at this point to carry out a harmonic expansion of our
variables introducing scalar harmonics $Q^{(k)}$ which satisfy the
Helmholtz equation $\3\nabla^2 Q^{(k)}=-k^2/\ell^2Q^{(k)}$, where $k$ is
a comoving (i.e. constant) eigenvalue. If $\nu$ is a non\hs
negative real wavenumber then in a flat $K=0$ universe, it is
associated with the  physical wavelengths $\lambda=2\pi \ell/
\nu$\footnote{A completely consistent theory of
CMB fluctuations in open models seems to be lacking until now
\cite{bi:panek}, and  even the issue of completeness of the set of
eigenfunctions of the  Laplacian operator seems unresolved.}
since when $K=0$, $k=\nu$, however if $K=-1$ then $k^2=\nu^2+1$
\cite{bi:LK}. Since observations tend to favor $\Omega_0\leq
1$\,\cite{bi:KT}, we  shall now restrict attention to the cases
$K=-1,0$. We consider the equations for $\Delta$ and $\wt{C}$, since
it is the latter variable (and not the GI curvature perturbation $C$)
that will be eventually conserved when $\Omega\not=1$. Writing equation
(\ref{eq:cdot}) in  terms of harmonic components, the evolution
equation for $\wt{C}^{(\nu)}$ becomes: \footnote{A brief summary of the
needed harmonic functions is given in Appendix B of \cite{bi:BDE}.}
%\newpage
\ber
\dot{\wt{C}}^{(\nu)}&=&-\case{4\ell^2H^3c^2_s}/{\left(1+w\right)}
\left(\case{\nu^2}/{\ell^2H^2}-\case{K}/{\ell^2H^2}\right)\Delta^{(\nu)}
+\case{4\ell^2H^3w}/{\left(1+w\right)}\left[\case{4K}/{\ell^2H^2}-
\case{\nu^2}/{\ell^2H^2}\right]\es^{(\nu)}\nonumber\\
&-&4\ell^3H^3\left(\case{\nu^2}/{\ell^2H^2}-\case{K}/{\ell^2H^2}\right)
\left[F^{(\nu)}+\Pi^{(\nu)}\right]\nonumber\\
&-& H^2\left[4K\ell-2\ell^3h\right]\left(\case{\nu^2}/{\ell^2H^2}-\case{K}/
{\ell^2H^2} \right)\Psi^{(\nu)}.
\label{eq:harmC2}
\ear
For large scales, i.e. wavelengths $\lambda\gg
H^{-1}\Rightarrow\case{\nu^2}/{\ell^2H^2}\ll
1$ this equation reduces to
\ber
\dot{\wt{C}}^{(\nu)}&=&+\case{4\ell^2H^3c^2_s}/{\left(1+w\right)}
\left(\case{K}/{\ell^2H^2}\right)\Delta^{(\nu)}
+\case{4\ell^2H^3w}/{\left(1+w\right)}\left[\case{4K}/{\ell^2H^2}
\right]\es^{(\nu)}\nonumber\\
%% FOLLOWING LINE CANNOT BE BROKEN BEFORE 80 CHAR
&+&4\ell^3H^3\left(\case{K}/{\ell^2H^2}\right)\left[F^{(\nu)}+\Pi^{(\nu)}\right]
\nonumber\\
%% FOLLOWING LINE CANNOT BE BROKEN BEFORE 80 CHAR
&+&H^2\left[4K\ell-2\ell^3h\right]\left(\case{K}/{\ell^2H^2}\right)\Psi^{(\nu)}.
\label{eq:harmC3}
\ear
Let us now examine the consequences of this equation for a number of
different cases.

If the background is flat, $K=0$, we immediately see that $\wt{C}=C$
{\em is conserved in the large scale limit even when entropy
perturbations and imperfect fluid  source terms are present}. This is
an important result, since we  can now use this conserved quantity to
write down a general solution  of the first order equation for
$\Delta^{(\nu)}$, (\ref{eq:deltadot}),  in the long wavelength limit:
\be
\Delta^{(\nu)}=\int^t_{t_0}\left[\int^t_{t_1}e^{A(t_2)}dt_2\right]
B^{(\nu)} (t_1)dt_1, \label{eq:int1}
\ee
where
\be
A(t)=3Hw-\case{\kappa h}/{2H},
\ee
\be
B^{(\nu)}(t)=\case{(1+w)}/{4\ell^2H}\tilde{C}^{(\nu)}+3\ell
\left(1+w\right)H\left[F^{(\nu)}+\Pi^{(\nu)}\right]\;,
\ee
and $t_0$ is the epoch at which an initial condition for
$\Delta^{(\nu)}$ is specified.

The conservation of $\wt{C}$ is also very important in the study of
perturbations in both standard and non\hs standard inflationary models  (based
on generalized gravity theories\footnote{The fluid flow approach to
perturbations can also be used to study perturbations in
generalized gravity theories. The basic idea is to treat all
additional contributions to the field equations except the Einstein
tensor part as contributions to an effective energy momentum tensor.
Effective fluid quantities are then easy to compute and GI quantities
based on spatial gradients can be defined as usual (see
\cite{bi:hwang}).}) since it can be used to directly connect the
amplitude of present day large scale structure, which came inside the
horizon during the matter  dominated era, to the initial conditions
just after horizon\hs  crossing during the inflationary era.

We now turn to the case when the background is an open ($K=-1$) model.
In this case $\wt{C}$ is conserved if the following inequalities hold:
\be
\frac{\nu^2}{\ell^2H^2}\ll 1\;~~and~~\frac{1}{\ell^2H^2}\ll 1,
\ee
i.e. if the long wavelength limit is valid up to $\nu=1$, provided
that the entropy perturbations and some of the imperfect fluid
terms do not exceed $\Delta^{(\nu)}$. However, in an open universe the
cosmological density parameter $\Omega$ is given  by
\be
\Omega=1-\frac{1}{\ell^2H^2},
\ee
so the second inequality only holds if $\Omega$ is close enough to
$1$, but then $C$ is also approximately conserved.
%%%%%%%%%%%%%%%%%%%%%%%%%%%%%%%%%%%%%%%%%%%
\section{Scalar fields} \label{sec:scalar}
%%%%%%%%%%%%%%%%%%%%%%%%%%%%%%%%%%%%%%%%%%%
Let us now extend the analysis carried out up to now to
a universe dominated by a scalar field $\phi$, as  is the case
during an inflationary phase; here we give only a brief  outline of
the method  (details can be found in \cite{bi:BED}).
We assume for simplicity that $\phi$ is minimally coupled (the
non\hs minimally coupled case will be considered elsewhere
\cite{bi:DB2}): then the Lagrangian density of $\phi$ is
\be
\ls_{\phi}=-\sqrt{-g}[\sfrac12 \phi_{,a}\phi^{,a} +V(\phi)]\;, \label{eq:lag}
\ee
where $V(\phi)$ is a general unspecified  potential assumed to be
responsible for inflation. We shall also assume that $\phi_{,a}$ is a
time\hs like vector\,\cite{bi:madsen}, and we make the  choice
\be
u^a=-\dot{\phi}^{-1}\phi^{,a}\;.
\ee
The former assumption will be well justified by a continuity argument
in an almost  FLRW universe, as it clearly holds in an exact FLRW
model; the latter choice of $u^a$ is uniquely determined by
$\phi^{,a}$ on requiring $u^au_a=-1$ (note that $\omega_{ab}=0$ with
this choice). Then the  energy\hs momentum tensor of $\phi$ takes the
perfect\hs fluid form (\ref{eq:emt}), with $p=\sfrac12 \dot{\phi}^2
-V(\phi)$ and $\mu=\sfrac12 \dot{\phi}^2 +V(\phi)$. With our choice of
$u^a$ the Klein\hs Gordon equation  $\Box\phi -V'(\phi)=0$ for $\phi$
(the prime indicates the derivative with respect to $\phi$) takes  the
form
\be
\ddot{\phi} +3H\dot{\phi} +V'(\phi)=0 \label{eq:KG}
\ee
in any space\hs time, while the spatial gradient of $\phi$ identically
vanishes: $h_a{}^b\phi_{,b}=0$. A gauge\hs invariant variable
associated with  $\phi$ is $h_a{}^b\dot{\phi}_{,b}$; from this we can
define a  dimensionless gauge\hs invariant scalar variable $\Phi$
simply related to the density perturbation $\Delta$:
\be
\Phi \equiv \ell h^{ac}(\ell \dot{\phi}^{-1} h_a{}^b \dot{\phi}_{,b})_{;c}\;
{}~~\Rightarrow ~~\Delta=\gamma\Phi \;.
\ee

It is then easy to show that equation (\ref{eq:deltadot}) for $\Delta$ still
holds, whereas the equation for $\wt{C}$ contains a further term
which can be interpreted as due to an entropy
perturbation\,\cite{bi:BED,bi:BST}.

The equation for the harmonic component $\wt{C}^{(\nu)}$ of $\wt{C}$
is then
\be
\dot{\wt{C}}^{(\nu)}= -\frac{4H^3 \ell^2}{\gamma}\left[\frac{\nu^2}{H^2\ell^2}
+\frac{6|K|}{H^2\ell^2}\left(\frac{\ddot{\phi}}{3H\dot{\phi}}
+\frac{7}{6}\right)\right]\Delta^{(\nu)}\;,
\label{eq:cndot}
\ee
while (\ref{eq:deltadot}) still holds for the components $\Delta^{(\nu)}$
and $\wt{C}^{(\nu)}$. Note that we left an explicit $|K|$ factor in
the above equation to show where $K$ appears: when $K=0$, $\wt{C}$
(\ref{eq:cdef}) reduces to $C$ and  the last term in (\ref{eq:cndot})
containing $|K|$ can be dropped, i.e. scalar field and barotropic
fluid perturbations satisfy formally  similar equations
[(\ref{eq:cndot}) reduces to (\ref{eq:cdot}) on putting $c_s^2=1$;
however this is only a formal identification: $c_s^2=c_s^2 (t)\not=1$
for a scalar field].
When $K=-1$ instead, the second term in the square brackets in
(\ref{eq:cndot}) will not be negligible in general, even for scales
$\lambda\gg H^{-1}$: $\wt{C}$ will then decrease if $V'>0$, but it
will increase in the opposite case.
We can ask now in which cases $\wt{C}^{(\nu)}$ will be conserved for
scales far outside the effective horizon $H^{-1}$ when $K=-1$.
It is clear from the above equation that in general $\wt{C}$ will not
be conserved, even if $\nu^2/(\ell H)^2\ll 1$, as the other term in the
parenthesis will be not negligible in general. However let us consider
the horizon crossing condition
\be
\nu/(\ell H)\approx 1 \Leftrightarrow\nu\approx
(1-\Omega_C)^{-\frac{1}{2}}> 1\;, \label{eq:hc}
\ee
where $\Omega_C$ is the value of $\Omega$ when the comoving scale
fixed by $\nu$ crosses the horizon; we see from the above relations
that scales with $\nu<1$ never cross the horizon.
Scales which are at present inside
the horizon, i.e. $\lambda<H_0^{-1}$, correspond to
$\nu>\nu_0\approx (1-\Omega_0)^{-\frac{1}{2}}$
 and therefore left the effective horizon when
$1-\Omega_C<1-\Omega_0$.
In standard inflationary models that predict $|1-\Omega_0|$ to
be exponentially close to zero all scales of major interest left the
horizon at the very end of inflation, and the corresponding
wavenumbers $\nu$  are exponentially large;  if instead
$|1-\Omega_0| \sim 1$ (i.e., if we have a minimal inflation),
than we can see many of the scales that left the horizon
from the very beginning of the inflationary phase, and in this case
$\nu\geq 1$. In any case these scales, and those outside the present
horizon ($\lambda>H_0^{-1}$) that lie in the interval
$1<\nu<(1-\Omega_0)^{-\frac{1}{2}}$ (practically vanishing for minimal
inflationary models), were far outside the horizon
only when $\Omega$ was already close enough to unity,
\be
\lambda\gg H^{-1} \Leftrightarrow \nu(1-\Omega)^{\frac{1}{2}}\ll 1
\Leftrightarrow 1-\Omega\ll 1-\Omega_C\;,
\ee
 in the last stage of inflation.
It follows that, in a universe which is fairly open today,
$\wt{C}^{(\nu)}$ cannot be conserved for these scales
in epochs when $|1-\Omega|\sim 1$.
%We note that this condition is satisfied for all the scales inside the
%present horizon if $|1-\Omega_0|\ll 1$,
% if $1-\Omega_0\ll 1$
%and at the onset of inflation if $\Omega$ is small.
%Then
For these scales  $\wt{C}^{(\nu)}$ will be practically conserved
whenever  $\Omega$ is very close to unity, provided also
\be
\frac{\ddot{\phi}}{3H\dot{\phi}} +\frac{7}{6} \sim 1\;,
\label{eq:cond1}
\ee
e.g. during a slow rolling phase implying $\ddot{\phi}\ll 3 H\dot{\phi}$.

Finally, those scales larger than the present horizon $H_0^{-1}$
corresponding to
$\nu<1$ have always been outside the horizon, so the
condition $\nu^2/(\ell H)^2\ll 1$ can be satisfied even when
$|1-\Omega|\sim 1$; however at these scales
 $\wt{C}^{(\nu)}$  will be not conserved in general, unless
\be
\left| \frac{1}{6}-\frac{V'}{3 H\dot{\phi}}\right| \ll 1\;.
\ee
One is mainly  concerned  with the  evolution of cosmological perturbations
which are at most
of the size of the present Hubble radius or slightly larger:
 if $|1-\Omega_0|\sim 1$
perturbations on such scales cannot be far larger than the effective
horizon from the moment they exit the horizon itself up to when $\Omega$
is close enough to unity. During such an epoch a quantity such as $\wt{C}$
cannot be conserved on these scales,  and the density  perturbation
evolution equation (\ref{eq:deltadot}) is no longer an independent equation
with a constant source term responsible for the growth of the density
perturbation itself. In the next section we will construct general conserved
quantities valid for particular inflationary models.
%%%%%%%%%%%%%%%%%%%%%%%%%%%%%%%%%%%%%%%%%%%
\section{Conserved quantities} \label{sec:gencons}
%%%%%%%%%%%%%%%%%%%%%%%%%%%%%%%%%%%%%%%%%%%
In the previous section  we discussed the evolution of the scalar
curvature variables $C$ and $\wt{C}$ and showed that for a scalar field neither
of them are conserved in general when $K\neq 0$ for scales much larger than the
Hubble radius. It is interesting to ask
whether it is possible to construct a quantity $\kk$ from the set of
perturbation variables that is conserved at any scale, even  for $K\neq 0$. It
would therefore generalize the two curvature variables already discussed.

Let's start by defining a general scalar quantity that generalizes
$C$ and $\wt{C}$: we found convenient to choose that it take the form
\be
\kk=-\case{4}/{3}\ell^2\Theta\zz+2\mu\ell^2\left(1-\case{2K}/{\ell^2\left(\mu+p
\right)}\right)\Delta+\ep X\zz+\ep Y\Delta,
\label{eq:gencondef}
\ee
where $X=X(t)$ and $Y=Y(t)$ are two functions to be determined, and
$\ep=0,1$. For
$\ep=0$ $\kk$ reduces to $\wt{C}$, and it reduces to $C$ if in
addition we choose  $K=0$. As we are now going to show, for $\ep=1$,
 there are
cases (clarified below) in which we can determine $X$ and $Y$ so that
$\kk$ is {\it conserved at any scale} (i.e. any mode $\kk^{(\nu)}$ is
conserved), both for a flat or for an open
universe (the following equations hold only for $K=0,-1$).
To find an evolution equation for $\kk$ , we take its time derivative
and use the equations for $\zz$ (\ref{eq:scalarZ}) and $\Delta$
(\ref{eq:deltadot}). Writing in terms of harmonic components, and keeping the
imperfect fluid source terms for generality, we obtain:
\ber
\dot{\kk}^{(\nu)} &=& \left\{ \ep \left[
\dot{Y}+3H\left(\gamma-1\right)Y-\case{1}/{2}\kappa \mu X+
\case{H^2c^2_s}/{\gamma}\left(\case{\nu^2}/{\ell^2H^2}-\case{4K}/{\ell^2 H^2}
\right)X\right] \right.\nonumber\\
&-&
\left.\case{4\ell^2H^3c^2_s}/{\gamma}\left(\case{\nu^2}/{\ell^2H^2}-
\case{K}/{\ell^2H^2}
\right) \right\} \Delta^{(\nu)}
+\ep \left[\dot{X}-2HX- \gamma Y\right]\zz^{(\nu)}\nonumber\\
&+& \case{H^2\left(\gamma-1\right)}/{\gamma}\left(
\case{\nu^2}/{\ell^2H^2}-\case{4K}/{\ell^2H^2}\right)
\left[\ep X-4H\ell^2\right]\es^{(\nu)}\nonumber\\
&+& \left\{ \ep \left[3 \ell\gamma H Y+\case{3}/{2} h X+\ell
H^2\left(\case{\nu^2}/{\ell^2H^2}-\case{4K}/{\ell^2
H^2}\right) X\right]\right.\nonumber\\
&-& \left. 4\ell^3
H^3\left(\case{\nu^2}/{\ell^2H^2}-
\case{K}/{\ell^2H^2}\right)\right\}\left[F^{(\nu)}
+\Pi^{(\nu)}\right] \nonumber\\
&+& \left[\ep \ell\gamma H^2 Y-4 K \ell H^2 +2
H^2\ell^3h\right]\left(\case{\nu^2}/{\ell^2H^2}-
\case{K}/{\ell^2H^2}\right)\Psi^{(\nu)}.
\label{eq:genconeq}
\ear
When $\ep=0$,\, $\kk^{(\nu)}=\wt{C}^{(\nu)}$ and this equation reduces to
equation (\ref{eq:harmC2}) of section \ref{sec:cons}.

It is clear from (\ref{eq:genconeq}) that $\kk^{(\nu)}$ could be
conserved  in general only if all the five coefficients of the equation
vanish, a condition that clearly cannot be satisfied by only two
arbitrary functions ($X$ and $Y$). However we are primarily interested
in the case of perfect fluids, for which $F=\Pi=\Psi=0$. Then for
such a fluid  either we take adiabatic perturbations ($\es=0$), or we
take it to  represent a minimally coupled scalar field
($F^{(\nu)}=\Pi^{(\nu)}= \Psi^{(\nu)}=0$), in which case
the entropy perturbation is
$p\es^{(\nu)}=\left(1-c^2_s\right)\mu\Delta^{(\nu)}$ \cite{bi:BED}. We
shall now focus our discussion on the latter case, for which
equation (\ref{eq:genconeq}) becomes:
\ber
\dot{\kk}^{(\nu)} &=& \left\{ \ep
\left[\dot{Y}+3H\left(\gamma-1\right)Y-\case{1}/{2}\kappa \mu X+
\case{H^2}/{\gamma}\left(\case{\nu^2}/{\ell^2H^2}-\case{4K}/{\ell^2H^2}\right)
X \right]\right. \nonumber\\
&-&\left. \case{4H^3\ell^2}/{\gamma}\left[ \case{\nu^2}/{\ell^2H^2}+
\case{K}/{\ell^2H^2}\left(3c^2_s-4\right)\right]
\right\}\Delta^{(\nu)}\nonumber\\
&+& \ep \left[\dot{X}-\gamma Y-2HX\right]\zz^{(\nu)}.
\label{eq:genconeq2}
\ear

Then the  condition for $\kk^{(\nu)}$ to be conserved is that the two
coefficients of $\Delta^{(\nu)}$ and $\zz^{(\nu)}$ vanish, i.e. only
the two arbitrary functions $X$ and $Y$ are needed; this justifies our
choice (\ref{eq:gencondef}). An equation similar to
(\ref{eq:genconeq2}) above is readily obtained from
(\ref{eq:genconeq}) for adiabatic perturbations of a perfect
(barotropic) fluid. In a more general case, such as a perfect fluid
with isocurvature perturbations ($\es\not=0$) or  an imperfect fluid
(as a non--minimally coupled scalar field \cite{bi:madsen}),
(\ref{eq:gencondef}) should be appropriately generalized with the
inclusion of further arbitrary functions.

{}From (\ref{eq:genconeq2}) with $\ep=1$, we obtain a pair of first order
differential equations for the variables $X$ and $Y$:
\ber
\lefteqn{
\dot{Y}+3H\left(\gamma-1\right)Y-\case{1}/{2}\kappa \mu X+\case{H^2}/{\gamma}
\left(\case{\nu^2}/{\ell^2H^2}-\case{4K}/{\ell^2H^2}\right)X}\nonumber\\
&&-\case{4H^3\ell^2}/{\gamma}\left[\case{\nu^2}/{\ell^2H^2}+
\case{K}/{\ell^2H^2}\left(3c^2_s-4\right)\right]=0
\label{eq:conY}
\ear
and
\be
\dot{X}-\gamma Y-2HX=0.
\label{eq:conX}
\ee
It must be noted that $X=X(\nu,t)$ and $Y=Y(\nu,t)$, i.e. both these
functions  depend on the wavenumber  $\nu$, as well as $t$:
for each $\nu$ the two equations above determine
a pair $X$, $Y$ such
that the corresponding harmonic component $\kk^{(\nu)}$ of $\kk$
will be conserved.
These  two equations
 can be combined to give a second order equation for $X$.
Substituting for $\gamma$ and $\mu$ in terms of $\dot{\phi}$
and $\ddot{\phi}$, we obtain:
\ber
\ddot{X}-\left[5H+\case{2\ddot{\phi}}/{\dot{\phi}}\right]\dot{X}
&+& \left[\case{1}/{2}
\dot{\phi}^2+6H^2+\case{4H\ddot{\phi}}/{\dot{\phi}}+
\case{1}/{\ell^2}\left(\nu^2-6K\right)
\right]X \nonumber \label{eq:Xev} \\
&=&
-24H\left[\case{K\ddot{\phi}}/{3H\dot{\phi}}
-\case{1}/{6}(\nu^2-7K)\right].
\label{eq:seccon} \ear
We will now apply the above equations to the following two cases
commonly used when discussing the evolution of energy\hs density
perturbations during an inflationary period: a coasting era,
characterized by a constant value of the density parameter $\Omega$
and De Sitter exponential inflation (see \cite{bi:EM1,bi:BED} for
a discussion of these models).
%%%%%%%%%%%%%%%%%%%%%%%%%%%%%%%%%%%%%%%%%%%
\section{Applications}
%%%%%%%%%%%%%%%%%%%%%%%%%%%%%%%%%%%%%%%%%%%
\subsection{The coasting solution: $\ell(t)=At$}
%%%%%%%%%%%%%%%%%%%%%%%%%%%%%%%%%%%%%%%%%%%
Using the results given in \cite{bi:EM1}, equation (\ref{eq:seccon})
becomes:
\be
\ddot{X}-\case{3}/{t}\dot{X}+\case{1}/{t^2}\left[3
+\case{1}/{A^2}\left(\nu^2-5K\right)\right]X
=4H\left(\nu^2-5K\right)\;,
\ee
which we can immediately integrate, giving, as in  \cite{bi:BED},
three classes of solution:
\paragraph{{\it a)} $\nu^2< A^2+5K$}\,:
\be
X(t)=X^At^{2+Q}+X^Bt^{2-Q}+4A^2t\;,
\ee
where $Q=\sqrt{\case{5K-\nu^2}/{A^2}+1}$ as before and $X^A$, $X^B$ are two
integration constants. The solution for $Y$ then follows from equation
(\ref{eq:conY}):
\be
Y(t)=\case{3}/{2}Q\left[X^At^{1+Q}-X^Bt^{1-Q}\right]-6A^2.
\ee
\paragraph{{\it b)} $\nu^2=A^2+5K$}\, ; in this case we obtain:
\be
X(t)=t^2\left[X^A+X^B\ln(t)\right]+4A^2t\;,
\ee
and
\be
Y(t)=\case{3}/{2}X^Bt-6A^2\;.
\ee
\paragraph{{\it c)} $\nu^2>A^2+5K$}\, ; the third case
corresponds to damped oscillations:
\be
X(t)=t^2\left\{X^A\cos\left[Q\ln(t)\right]+X^B\sin\left[Q\ln(t)\right]\right\}
+4A^2t\;,
\ee
and
\be
%% FOLLOWING LINE CANNOT BE BROKEN BEFORE 80 CHAR
Y(t)=-\case{3}/{2}Qt\left\{X^A\sin\left[Q\ln(t)\right]-X^B\cos\left[Q\ln(t)\right]
\right\}-6A^2\;.
\ee
Substituting these results into equation (\ref{eq:gencondef}) we can obtain an
expression for the conserved quantity $\kk$.
%%%%%%%%%%%%%%%%%%%%%%%%%%%%%%%%%%%%%%%%%%%
\subsection{De Sitter exponential expansion: $\ell(t)=A\exp{\omega t}$}
%%%%%%%%%%%%%%%%%%%%%%%%%%%%%%%%%%%%%%%%%%%
This time the second order differential equation for
$X(t)$ (\ref{eq:seccon}) becomes:
\be
\ddot{X}-3w\dot{X}+\left[2w^2+\case{1}/{A^2}\left(\nu^2-5K\right)
\exp(-2wt)\right]X=4w\left(\nu^2-5K\right)w\;.
\ee
Again, as in the previous example we can integrate this equation to
give three classes of solution:
\paragraph{{\it a)} $\nu^2-5K< 0$}\,:
\be
X(t)=\case{1}/{C^2p}\left[X^A\exp(2C\sqrt{p})+X^B\exp(-2C\sqrt{p})\right]
+\case{C^2}/{w}\case{(\nu^2-5K)}/{p}\;,
\ee
where $C=\sqrt{\case{\nu^2-5K}/{4w^2A^2}}$, $p=\exp(-2wt)$ and $X^A$,
$X^B$ are two integrating constants. The corresponding solution for $Y$
is:
\ber
Y(t)&=&\case{1}/{\gamma}\left[\case{2\omega}/{C^2p^2}\left\{X^A\left((1-p)
\exp(2C\sqrt{p})-\case{Cp}/{\sqrt{p}}\exp(-2C\sqrt{p})\right)\right.\right.
\nonumber\\
&+&\left. X^B\left((1-p)\exp(-2C\sqrt{p})+\case{Cp}/{\sqrt{p}}\exp(2C\sqrt{p})
\right)\right\}\nonumber\\
&+& \left.\case{2C^2}/{p^2}(\nu^2-5K)(1-p)\right]\;.
\ear
\paragraph{{\it b)} $\nu^2-5K=0$}\,; in this case we obtain:
\be
X(t)=X^A\exp(-\case{1}/{2}p)+X^B\exp(-p)\;,
\ee
and
\be
Y(t)=\case{w}/{\gamma}\left[X^A(p-2)\exp(-\case{1}/{2}p)+2X^B(p-1)\exp(-p)
\right]\;.
\ee
\paragraph{{\it c)} $\nu^2-5K>0$}\,:
\be
X(t)=\case{1}/{C^2p}\left[X^A\cos(2C\sqrt{p})+X^B\sin(2C\sqrt{p})\right]
+\case{C^2}/{w}\case{(\nu^2-5K)}/{p}\;,
\ee
and
\ber
Y(t)&=&\case{1}/{\gamma}\left[\case{2\omega}/{C^2p^2}\left\{X^A\left((1-p)
\cos(2C\sqrt{p})+\case{Cp}/{\sqrt{p}}\sin(2C\sqrt{p})\right)\right.\right.
\nonumber\\
&+& \left. X^B\left((1-p)\sin(2C\sqrt{p})-\case{Cp}/{\sqrt{p}}\cos(2C\sqrt{p})
\right)\right\}\nonumber\\
&+& \left.\case{2C^2}/{p^2}(\nu^2-5K)(1-p)\right]\;.
\ear
Substituting for $X(t)$ and $Y(t)$ into equation
(\ref{eq:gencondef}) we can obtain an expression for the
corresponding  conserved quantity $\kk$.
%%%%%%%%%%%%%%%%%%%%%%%%%%%%%%%%%%%%%%%%%%%
\section{Transfer functions} \label{sec:transfer}
%%%%%%%%%%%%%%%%%%%%%%%%%%%%%%%%%%%%%%%%%%%
In the previous two sections we have illustrated a method to obtain a
conserved quantity $\kk^{(\nu)}$ via two auxiliary functions
$X(\nu,t)$  and $Y(\nu,t)$. Without going into further details, in the
following we shall sketch how to use $\kk^\inu$ to connect the
amplitude of  perturbations at different epochs.

Typically, one is interested in connecting an early universe
(inflationary) era with the present matter dominated era.  For this
reason, and for the sake of simplicity, we will consider eras in which
we either have a perfect fluid with adiabatic perturbations, or a
minimally coupled scalar field. In both these cases, the
evolution of perturbations is given by a system of equations like
those of section \ref{sec:cons}, i.e.
\ber
\dot{\Delta}^\inu & = & a(t) \Delta^\inu +b(t) \zz^\inu\;,
\label{eq:devol} \\
\dot{\zz}^\inu & = & c(t) \Delta^\inu + d(t) \zz^\inu \;,
\ear
where at least one of the functions $a,b,c,d$ depends on $\nu$ [we
remind the reader that the exception is the dust case, for which
$\wt{C}$ is  conserved at all scales, even in an open universe, see
(\ref{eq:harmC2})], and all depend on the equation of state valid in
the given era.

We can write equation (\ref{eq:gencondef}) in the form
\be
\kk_\inu=\kappa_1(\nu,t)\Delta^\inu +\kappa_2(\nu,t)\zz^\inu\;,
\label{eq:kkdef}
\ee
where $\kappa_1$ and $\kappa_2$ depend on $X(t)$ and $Y(t)$. Using
(\ref{eq:kkdef}) we then obtain an equation of the same form as
(\ref{eq:genconeq2}), and requiring the vanishing of the coefficients
of this equation we get (\ref{eq:conY}) and (\ref{eq:conX}), which can
then be integrated (at least in some cases) to give us the functions
$X$ and $Y$. With these results in mind, while at the same time
using  (\ref{eq:kkdef}) to substitute for $\zz^\inu$ in
(\ref{eq:devol}), we obtain a first order equation for $\Delta^\inu$:
\be \dot{\Delta}^\inu=\wt{a}(t) \Delta^\inu +\wt{b}(t)\zz^\inu\;,
\label{eq:newd} \ee
and since $\kk^\inu=const$, this equation can always be integrated
(at least in principle). Then, neglecting a decaying mode, this will
give
\be
\Delta^\inu(t)=F(\nu,t)\kk^\inu \;,
\ee
or an integral of the form (\ref{eq:int1}) in the general case.
If we want to connect two different eras with different equation of
state, we have to repeat the procedure sketched above for the two
different cases. Then using $\kk^\inu (t_1)=\kk^\inu (t_2)$ one
obtains
\be
\Delta^\inu (t_1) = T(\nu,t_1,t_2) \Delta^\inu (t_1)\;,
\ee
where $T(\nu,t_1,t_2)=F(\nu,t_2)/ F(\nu,t_1)$ is the required
transfer function.

Thus we have obtained an algorithm through which, provided that the
two auxiliary functions $X$ and $Y$ can be explicitly determined, the
evolution of the density perturbation $\Delta^\inu$ s given, for any
scale fixed by $\nu$, by a first order equation with the conserved
quantity $\kk^\inu$ as source term. Using this, a transfer function is
readily obtained, thus generalizing the procedure illustrate for
example in  \cite{bi:lily,bi:efst} to any $\nu$ and $K=0,-1$.
%%%%%%%%%%%%%%%%%%%%%%%%%%%%%%%%%%%%%%%%%%%
\section{Discussion}
%%%%%%%%%%%%%%%%%%%%%%%%%%%%%%%%%%%%%%%%%%%
In the present paper we discussed under what conditions the curvature
variables $C$ and $\wt{C}$ are conserved both in the case of a general
imperfect
fluid and in detail for a minimally coupled scalar field. It was shown
that in the general case $\wt{C}$ is only conserved for large scales
when the background is taken to be flat or almost flat (i.e.
$\Omega\simeq  1$), while for a minimally coupled scalar
field an additional condition (\ref{eq:cond1}) must be satisfied in the
almost flat case. We then
explored the possibility of defining a gauge\hs invariant curvature
variable  $\kk$ which would be conserved in a  more general set of
circumstances for all scales of
interest. We showed that at least for some models this possibility can indeed
be realized and illustrated this with the example of a minimally
coupled scalar field evolving either in a coasting phase
($\Omega=const$) or during a De Sitter inflationary era. We also
discussed how to use $\kk$ to derive a transfer function in order to
connect the amplitude of perturbations at different epochs.

Although we restricted our attention to rather special examples, all the
equations have been derived in general, so this prescription can be applied to
a wider variety of models both inflationary and non\hs inflationary.
In particular, applications to generalized gravity theories will be of great
interest in view of the recent growing activity in such theories and the
generation and evolution of perturbations during a possible inflationary phase
and their final spectrum at second horizon crossing time deserve special
attention. These issues will be discussed in a future paper \cite{bi:DB2}.
%%%%%%%%%%%%%%%%%%%%%%%%%%%%%%%%%%%%%%%%%%%
\section*{Acknowledgments}
%%%%%%%%%%%%%%%%%%%%%%%%%%%%%%%%%%%%%%%%%%%
PKSD thanks George Ellis for useful discussions, the School of
Mathematical Sciences, QMW for hospitality while some of this work
was carried out and the FRD (South Africa) for financial support. MB
thanks SERC (UK) and MURST (Italy) for  financial support.
%%%%%%%%%%%%%%%%%%%%%%%%%%%%%%%%%%%%%%%%%%%

\end{document}